\documentclass[twoside, epsfig]{article}

\input oejv2.sty

\begin{document}

\OEJVhead{12 2022}
\OEJVtitle{ASASSN-18aan revisited}
\OEJVauth{Nesci, Roberto$^1$; Vagnozzi, Antonio$^2$; Valentini, Stefano$^2$} 
\OEJVinst{INAF/IAPS, Via Fosso del Cavaliere 100, I-00133, Roma, Italy {\tt \href{mailto:roberto.nesci@inaf.it}{roberto.nesci@inaf.it}}}

\OEJVinst{MPC589, Via S.Lucia n.68, I-05039, Stroncone (TR), Italy {\tt \href{mailto:a.vagnozzi@mpc589.com}{a.vagnozzi@mpc589.com}}}

\OEJVabstract{The light curve of the cataclismic variable ASASSN-18aan is studied using recent observations of the MC589 Observatory, giving an orbital Period and Epoch fully consistent with the data obtained after the discovery flare in 2018. Archival data from ASASSN, ZTF and Gaia were used to check if its flares have a quasi-periodic behaviour. A recurrency time scale of about 11 months is found, confirming a previous tentative result using the historic plate archive of the Asiago Observatory. The next outbursts are expected by April 2023 and March 2024.
}

\begintext

\section{Introduction}\label{secintro} 
 The transient ASASSN-18aan (00:46:08.03 +62:10:05.0) was discovered on 2018-11-29 by the ASASSN project \citep{shap14} in the V filter with a peak luminosity V=15.4. Its nature of cataclismic variable was confirmed by \citet{Nesci} with an optical spectrum showing large emission lines and an X-ray counterpart in the SWIFT-XRT archive data. A search in the Asiago and Heidelberg plate archives showed a few pre-discovery flares since 1967, giving a possible recurrence time scale of about 11 months. 
Timely optical follow-up of the 2018 flare (e.g. T. Vanmunster, vsnet-alert 22806, Tordai and Kato, vsnet-chat 22816, Dufoer et al., vsnet-alert 22821, Wakamatsu et al. vsnet-alert 22825) discovered its nature of eclipse binary with amplitude 0.45 mag, a period of 0.14941 days, and the presence of superhumps in the descending phase of the flare. 
 An extensive photometric and spectroscopic follow-up during the flare and in the following months was made by \citet{Wakamatsub} who derived physical properties of this rather peculiar binary system, and classified it as an SU UMa-type cataclismic variable. The light curve has two minima, one slightly more deep the the other; the average r-band magnitude in quiescence was 17.2 with peak to peak amplitude 0.4 mag, and the period was found 0.149454(3) days, confirmed by the radial velocity curve. 

Based on archive images, \citet{Nesci} made a forecast for the next flares around  2019-10-30, 2020-09-30 and 2021-08-30. To check if the recurrence time scale of the flares was correct, we made an archival search in the ASASSN and in the ZTF  \citep{Masci} databases for detections of this star. Optical follow-up was also started since January 2023 with the 30 cm telescope of the S. Lucia di Stroncone Observatory (MPC589), to look for the expected outburst in Spring 2023, and to check a possible period variation since 2018 during quiescence.

\section{Archival Observations }

We downloaded from the ASASSN website the observations of ASASSN-18aan since 2018-09-09 (HJD 2458370.89). Given its faint quiescent level (g=17.5 in PanSTARRS DR1 \citep{Flew}) the star is generally not detected in quiescence by the ASASSN survey, but flaring states are expected to be detected even in full Moon nights, given that they are about g=15.5. Overall 5 high states are present in this database.
A further flaring state was detected by Gaia on 2020-09-01 (Gaia alert Gaia20ecl, AT2020sls \citep{Gaia}). Two further high states are present in the ZTF DR14. 

Rather surprisingly, this star is not present in the ZTF alert service, despite it is well visible in all the public images of the ZTF archive at IRSA \citep{Masci}. The public lightcurve in the ZTF DR14 database covers the time interval 2010-08-02 to 2019-11-03, and contains two further high states of the source. However, not all the relevant ZTF images are present in this lightcurve. Given that all the images are available at IRSA shortly after they are secured, and can be freely downloaded in 16 bit FITS format, we downloaded the ZTF images around the ASASSN and Gaia flares down to the present date, to check the reality of the flares, as well as their amplitude and duration. Photometry of the ZTF images was made with IRAF/apphot using the PanSTARRS DR1 catalog for the comparison stars.

Table 1 collects the dates of the flares, the MJD, the telescope, the distance from the "reliable" preceding flare, the expected date from \citet{Nesci}, and a comment.

\begin{table} 
\caption{High states of ASASSN-18aan from archive data.}\vspace{3mm}  
\centering
\begin{tabular}{lcrrrr}
\hline
MJD & YYYY-MM-DD& Tel& DeltaT & Expected& Comment\\
\hline
58451& 2018-11-29& ASASSN&   0&           &Detection\\
58457& 2018-12-05& ZTF   &   6&           & several points \\
58739& 2019-09-13& ZTF   & 288& 2019-10-30& several points \\
59065 &2020-08-04& ASASSN& 326&           & ZTF quiescence\\
59093 &2020-09-01& Gaia  & 354& 2020-09-01& Gaia20ecl \\
59223 &2021-01-09& ASASSN& 130&           & ZTF quiescence\\
59431 &2021-08-05& ASASSN& 338& 2021-08-30& several points\\
59740 &2022-06-09& ASASSN& 309&           &several points\\
\hline
\end{tabular}\label{tab1}
\end{table}

In the following we give a short description of the flares as derived from the available data.
\begin{itemize}
\item 58451: the flare is well followed by ASASSN from 58434 to 58486, by ZTF from 58444 to 58495, and by AAVSO observers from 58457 to 58487. Extensive coverage was made by \citet{Wakamatsub} from 58463 to 58517. All these observations show an intranight variability period of 3.5 hours with 0.45 amplitude present during the flare, and also in the subsequent quiescent phase.

\item 58739: detected only by ZTF from 58736 to 58745; peak rmag=15.7 on 58739, back to rmag=16.7 on 58746; flare duration about one week.

\item 59065: a single point in ASASSN with Full Moon, with nearby points flagged as upper limits. The dense coverage by ZTF shows the source always in a quiescent state, so it was a false alarm.

\item 59093: a single point from Gaia alert Gaia20ecl; no data available around that date from ASASSN. From ZTF images the high state is confirmed on MJD 59093 but the star was again at the normal level on 59097, so this flare was quite short.

\item 59223: a single point from ASASSN. ZTF images around the date show the source in quiescence, so we judge this point as unreliable.

\item 59431 Several good points are present in ASASSN, from 59430 to 59433; ZTF images confirm the high state on 59428 at r=15.91 but sees the source quiescent already on 59435, so this flare is shorter than a week.

\item 59740 : three good points around g=15.6 are present in ASASSN; the following ASASSN observations are 10 days later and are upper limits only. Unfortunately the ZTF has no data in this period.
\end{itemize}

From Table 1, excluding the ASASSN unreliable detections, we derive an average recurrence time of 323 days (10.6 months), confirming the previous estimate of about 11 months by \citet{Nesci} from Asiago archive plates. This recurrence time interval however is not constant but shows fluctuations of about one month.

About the flares duration, in the 2018 event the high state lasted 15 days, and the star returned to the quiescent level in 6 days. On the contrary, the following flares were rather short, less than 5 days,
so that bad weather conditions can easily prevent the detection and/or a good coverage of this kind of flares in the source.

The outburst on MJD 58451 was quite longer than the other ones, and may be called a superoutburst.  Further observations along several years will be needed to understand if such superoutbursts have a recurrence time scale. Starting from the observed June 2022 flare, the next flares are expected around the end of April 2023 (2023-04-29), when the star will be observable before the morning astronomical twilight, and mid March 2024 (2024-03-15), with the star observable soon after sunset.

\section{ Our intranight observations}
To check the constancy of the period we started an observation campaign on January 2023 using the 30 cm F/10 Schmidt Cassegrain telescope of the MPC589 Observatory, equipped with a CCD camera with a KAF1001E sensor: pixel size is 24 micron and FOV 29 arcmin. Given the faintness of the star no filter was used.  Exposure time was set at 240 s; flat field and dark current corrections were applied in a standard way. The log of the  observations is reported in Table \ref{tab2}.

\begin{table} 
\caption{Observation log. Columns $N$ gives the number of usable exposures.}\vspace{3mm}  
\centering
\begin{tabular}{lcr}
\hline
	 Night & Time-span [hours] & $N$   \\ \hline \hline
	 29 January, 2023& \hspace{1.8mm}5.8 	& 87  \\
	 11 February, 2023 & \hspace{1.8mm}4.6 	& 69 \\
  14 February, 2023 & \hspace{1.8mm}4.2	& 63  \\
  17 February, 2023 & \hspace{1.8mm}4.1 	& 61 \\
  20 February, 2023 & \hspace{1.8mm}2.7 	& 40 \\
\hline
\end{tabular}\label{tab2}
\end{table}




\begin{table}
\caption{Comparison stars for the MPC589 observations. Coordinates and Magnitudes from Gaia DR3. x is the variable ASASSN-18aan}\vspace{3mm}  
\centering
\begin{tabular}{llrr}
\hline
RAJ2000&DEJ2000&id&Gmag\\
\hline\hline
11.53346 &62.16804 &x&16.843\\
11.43663 &62.13594 &c&13.343\\
11.44659 &62.16367 &d&16.676\\
11.48484 &62.19037 &e&13.976\\
11.48976 &62.21424 &f&14.421\\
11.49427 &62.18062 &g&15.004\\
11.50503 &62.15483 &h&16.187\\
11.51641 &62.12781 &i&15.388\\
11.53435 &62.13671 &k&15.370\\
11.54303 &62.13374 &l&15.313\\
11.57448 &62.14321 &m&13.778\\
11.59062 &62.21265 &n&14.470\\
11.60098 &62.16753 &o&14.419\\
11.60953 &62.20417 &q&15.302\\
11.61467 &62.11932 &r&14.162\\
11.61698 &62.13636 &s&15.514\\
11.63723 &62.18538 &t&13.950\\
11.65776 &62.14401 &u&14.424\\
\hline
\end{tabular}\label{tab3}
\end{table}



\section{Data analysis}\label{secdata}
 Seventeen comparison stars (see Table \ref{tab3}) were taken from the Gaia DR3 catalog, with BP-RP color index similar to that of the variable, adopting their G magnitudes as representative of our unfiltered ones. 
 
 We verified that the Gaia G magnitudes of the comparison stars were equal to the PanSTARRS r ones at a 0.02 mag level. 
 
 Aperture photometry was made with IRAF/apphot using a radius equal to the stellar FWHM. The photometric error was derived from the rms deviation of the comparison stars with respect to the linear fit of the instrumental vs catalog magnitudes: typical values were 0.04 mag.

 Our photometric accuracy around mag 17 was barely enough to distinguish the primary and secondary minima, so we forced our period search around the published one, supported by the radial velocity curve.

We computed the period of the light curve and the epoch of minimum with two methods, the Fast Fourier Trasform (FFT), and the String Length Minimization, using a code written by one of us (SV). The SLM is similar to the PDM method used by \citet{Wakamatsub} so that systematic differences are less likely: anyway both methods gave the same results: P$_{orb}$=0.149440(40) d and Epoch=2459993.309810 HJD. 
We checked our results using Period04 \citep{Lenz05}, for the FFT, and PDM as implemented in Peranso \citep{Paunzen16}, finding fully consistent results. The formal uncertainty on the period was given by Peranso, using the algorithm by \citet{Stel}.

The resulting phased light curve is shown in Fig. 1.

\begin{figure}[htbp]
\centering
\includegraphics[width=14cm]{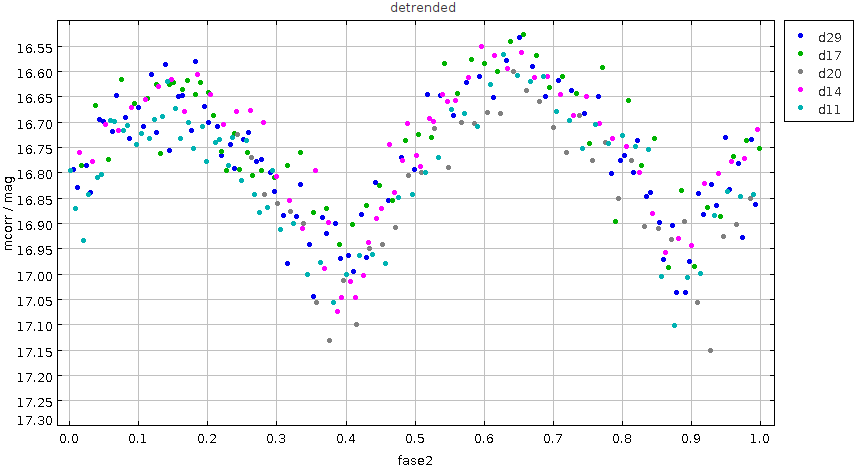} 
\caption{Phase reduced light curve of ASASSN-18aan from our 30cm telescope from the data of the five nights in Table \ref{tab2}.}
\label{fig1}
\end{figure}

A further check of the period was made comparing our Epoch of minimum, transformed in BJD with the code by \citet{xx} (2459993.310587), with that published by \citet{Wakamatsub}, rounding to the nearest integer the number of cicles: the period computed in this way resulted 0.149452 d, fully consistent with their 0.149454(3). We conclude that no period change happened since the end of the 2018 superoutburst, and we infer therefore that no significant mass and angular momentum was lost from the stellar system.
	
\section{Results}\label{secresults}

We performed an archival search for flares of the cataclismic variable ASASSN-18aan after its discovery outburst in november 2018, using the ASASN, the ZTF and the Gaia Alert archives.
A total of 4 further flares were detected, indicating an average recurrency time of 320 $\pm$30 days.
All the recorded flares were much shorter than the discovery one, 
suggesting that it could be considered as a superoutburst.
The next flares are therefore expected around the end of April 2023 and the end of March 2024.

We started a follow-up program of intranight observations at the MPC589 Observatory, confirming the period published by \citet{Wakamatsub} after the 2018 major flare, so that no significant mass loss happened to the system since then. It would be useful to make a new measure of the period after that the source will undergo another large flare.

\setcounter{secnumdepth}{0}
\OEJVacknowledgements{
We made extensive use of the ZTF public images archive at IRSA, and of the ASASSN public archive. We also used the ESA Gaia, DPAC and the Photometric Science Alerts Team.
	} \\


\end{document}